\documentclass{webofc}
\usepackage[varg]{txfonts}   % Web of Conferences font

\begin{document}
\title{Systematic effects on the upcoming NIKA2 LPSZ scaling relation}

\author{%
  \lastname{A.~Moyer-Anin}\inst{\ref{LPSC}}\fnsep\thanks{ alice.moyer@lpsc.in2p3.fr}
  \and  R.~Adam \inst{\ref{OCA}}
  \and  P.~Ade \inst{\ref{Cardiff}}
  \and  H.~Ajeddig \inst{\ref{CEA}}
  \and  P.~Andr\'e \inst{\ref{CEA}}
  \and  E.~Artis \inst{\ref{LPSC},\ref{Garching}}
  \and  H.~Aussel \inst{\ref{CEA}}
  \and  I.~Bartalucci \inst{\ref{Milan}}
  \and  A.~Beelen \inst{\ref{LAM}}
  \and  A.~Beno\^it \inst{\ref{Neel}}
  \and  S.~Berta \inst{\ref{IRAMF}}
  \and  L.~Bing \inst{\ref{LAM}}
  \and  O.~Bourrion \inst{\ref{LPSC}}
  \and  M.~Calvo \inst{\ref{Neel}}
  \and  A.~Catalano \inst{\ref{LPSC}}
  \and  M.~De~Petris \inst{\ref{Roma}}
  \and  F.-X.~D\'esert \inst{\ref{IPAG}}
  \and  S.~Doyle \inst{\ref{Cardiff}}
  \and  E.~F.~C.~Driessen \inst{\ref{IRAMF}}
  \and  G.~Ejlali \inst{\ref{Tehran}}
  \and  A.~Gomez \inst{\ref{CAB}} 
  \and  J.~Goupy \inst{\ref{Neel}}
  \and  C.~Hanser \inst{\ref{LPSC}}
  \and  S.~Katsioli \inst{\ref{Athens_obs}, \ref{Athens_univ}}
  \and  F.~K\'eruzor\'e \inst{\ref{Argonne}}
  \and  C.~Kramer \inst{\ref{IRAMF}}
  \and  B.~Ladjelate \inst{\ref{IRAME}} 
  \and  G.~Lagache \inst{\ref{LAM}}
  \and  S.~Leclercq \inst{\ref{IRAMF}}
  \and  J.-F.~Lestrade \inst{\ref{LERMA}}
  \and  J.~F.~Mac\'ias-P\'erez \inst{\ref{LPSC}}
  \and  S.~C.~Madden \inst{\ref{CEA}}
  \and  A.~Maury \inst{\ref{CEA}}
  \and  P.~Mauskopf \inst{\ref{Cardiff},\ref{Arizona}}
  \and  F.~Mayet \inst{\ref{LPSC}}
  \and  A.~Monfardini \inst{\ref{Neel}}
  \and  M.~Mu\~noz-Echeverr\'ia \inst{\ref{LPSC}}
  \and  A.~Paliwal \inst{\ref{Roma}}
  \and  L.~Perotto \inst{\ref{LPSC}}
  \and  G.~Pisano \inst{\ref{Roma}}
  \and  E.~Pointecouteau \inst{\ref{Toulouse}}
  \and  N.~Ponthieu \inst{\ref{IPAG}}
  \and  G.~W.~Pratt \inst{\ref{CEA}}
  \and  V.~Rev\'eret \inst{\ref{CEA}}
  \and  A.~J.~Rigby \inst{\ref{Leeds}}
  \and  A.~Ritacco \inst{\ref{INAF}, \ref{ENS}}
  \and  C.~Romero \inst{\ref{Pennsylvanie}}
  \and  H.~Roussel \inst{\ref{IAP}}
  \and  F.~Ruppin \inst{\ref{IP2I}}
  \and  K.~Schuster \inst{\ref{IRAMF}}
  \and  A.~Sievers \inst{\ref{IRAME}}
  \and  C.~Tucker \inst{\ref{Cardiff}}
  %\and  R.~Zylka \inst{\ref{IRAMF}}
}
\institute{
  Universit\'e Grenoble Alpes, CNRS, Grenoble INP, LPSC-IN2P3, 38000 Grenoble, France
  \label{LPSC}
  \and	
  Universit\'e C\^ote d'Azur, Observatoire de la C\^ote d'Azur, CNRS, Laboratoire Lagrange, France 
  \label{OCA}
  \and
  School of Physics and Astronomy, Cardiff University, CF24 3AA, UK
  \label{Cardiff}
  \and
  Universit\'e Paris-Saclay, Université Paris Cité, CEA, CNRS, AIM, 91191, Gif-sur-Yvette, France
  \label{CEA}
  \and
  Max Planck Institute for Extraterrestrial Physics, 85748 Garching, Germany
  \label{Garching}
  \and
  INAF, IASF-Milano, Via A. Corti 12, 20133 Milano, Italy.
  \label{Milan}
  \and
  Aix Marseille Univ, CNRS, CNES, LAM, Marseille, France
  \label{LAM}
  \and
  Universit\'e Grenoble Alpes, CNRS, Institut N\'eel, France
  \label{Neel}
  \and
  Institut de RadioAstronomie Millim\'etrique (IRAM), Grenoble, France
  \label{IRAMF}
  \and 
  Dipartimento di Fisica, Sapienza Universit\`a di Roma, I-00185 Roma, Italy
  \label{Roma}
  \and
  Univ. Grenoble Alpes, CNRS, IPAG, 38000 Grenoble, France
  \label{IPAG}
  \and
  Institute for Research in Fundamental Sciences (IPM), Larak Garden, 19395-5531 Tehran, Iran
  \label{Tehran}
  \and
  Centro de Astrobiolog\'ia (CSIC-INTA), Torrej\'on de Ardoz, 28850 Madrid, Spain
  \label{CAB}
  \and
  National Observatory of Athens, IAASARS, GR-15236, Athens, Greece
  \label{Athens_obs}
  \and
  Faculty of Physics, University of Athens, GR-15784 Zografos, Athens, Greece
  \label{Athens_univ}
  \and
  High Energy Physics Division, Argonne National Laboratory, Lemont, IL 60439, USA
  \label{Argonne}
  \and  
  Instituto de Radioastronom\'ia Milim\'etrica (IRAM), Granada, Spain
  \label{IRAME}
  \and
  LERMA, Observatoire de Paris, PSL Research Univ., CNRS, Sorbonne Univ., UPMC, 75014 Paris, France  
  \label{LERMA}
  \and
  School of Earth \& Space and Department of Physics, Arizona State University, AZ 85287, USA
  \label{Arizona}
  \and
  Univ. de Toulouse, UPS-OMP, CNRS, IRAP, 31028 Toulouse, France.
  \label{Toulouse}
  \and
  School of Physics and Astronomy, University of Leeds, Leeds LS2 9JT, UK
  \label{Leeds}
  \and
  INAF-Osservatorio Astronomico di Cagliari, 09047 Selargius, Italy
  \label{INAF}
  \and 
  LPENS, ENS, PSL Research Univ., CNRS, Sorbonne Univ., Universit\'e de Paris, 75005 Paris, France 
  \label{ENS}
  \and  
  Department of Physics and Astronomy, University of Pennsylvania, PA 19104, USA
  \label{Pennsylvanie}
  \and
  Institut d'Astrophysique de Paris, CNRS (UMR7095), 75014 Paris, France
  \label{IAP}
  \and
  University of Lyon, UCB Lyon 1, CNRS/IN2P3, IP2I, 69622 Villeurbanne, France
  \label{IP2I}
}

\abstract{%
In cluster cosmology, cluster masses are the main parameter of interest. They are needed to constrain cosmological parameters through the cluster number count. As the mass is not an observable, a scaling relation is needed to link cluster masses to the integrated Compton parameters Y, {\it i.e.} the Sunyaev-Zeldovich observable (SZ). Planck cosmological results obtained with cluster number counts are based on a scaling relation measured with clusters at low redshift ($z$<0.5) observed in SZ and X-ray.
In the SZ Large Program (LPSZ) of the NIKA2 collaboration, the scaling relation will be obtained with a sample of 38 clusters at intermediate to high redshift ($0.5<z<0.9$) and observed at high angular resolution in both SZ and X-ray.
Thanks to analytical simulation of LPSZ-like samples, we take into account the LPSZ selection function and correct for its effects. Besides, we show that white and correlated noises in the SZ maps do not affect the scaling relation estimation. 
}
\maketitle
\section{Introduction}
\label{intro}
Clusters of galaxies are used to constrain cosmological parameters through the cluster number count:  $\dfrac{{\rm d} N}{{\rm d} M{\rm d}z}$ \cite{cluster_nb_count}, 
where $N$ is the number of clusters, $M$ the cluster mass and $z$ the redshift. To have access to cluster masses, several methods exist using for example the weak and strong lensing effects \cite{lensing}
or multiwavelength observations combining X-ray and millimiter data.
In the millimiter domain clusters are observed with the Sunyaev-Zeldovich effect (SZ) \cite{SZ_effect}, which is the inverse Compton scattering of CMB photons on cluster ionized gas. It results in a shift of the CMB spectrum to high frequencies. The intensity is linked to the Compton parameter $y$, defined as : $y \propto \int P_e(r) {\rm d}l$ where $P_e(r)$ is the electronic pressure and $l$ is the line of sight.
This effect is redshift independent and has a characteristic spectral feature that makes possible cluster detections and observations up to high redshift.
Past and future large scale surveys, such as Planck \cite{Planck}, ACT \cite{ACT}, SPT \cite{SPT}, SO \cite{SO}, CMB-S4 \cite{CMBS4}, observe in the millimeter domain to get cluster catalogs. They contain information on the observed clusters, such as the redshift, the hydrostatic equilibrium mass $M_{500}$ and the SZ obervable $Y_{500}$. $M_{500}$ is defined as 
the mass enclosed in a sphere with density 500 times the critical density of the universe at the cluster's redshift. $Y_{500}$ is defined as the integral of the Compton parameter up to $R_{500}$.
For unresolved surveys, a $Y$-$M$ scaling relation (SR) that links the mass of the cluster to the SZ observable is needed. 
\section{Y-M scaling relation}
\label{sec-1}
\begin{figure}
    \centering
    \includegraphics[scale=0.20]{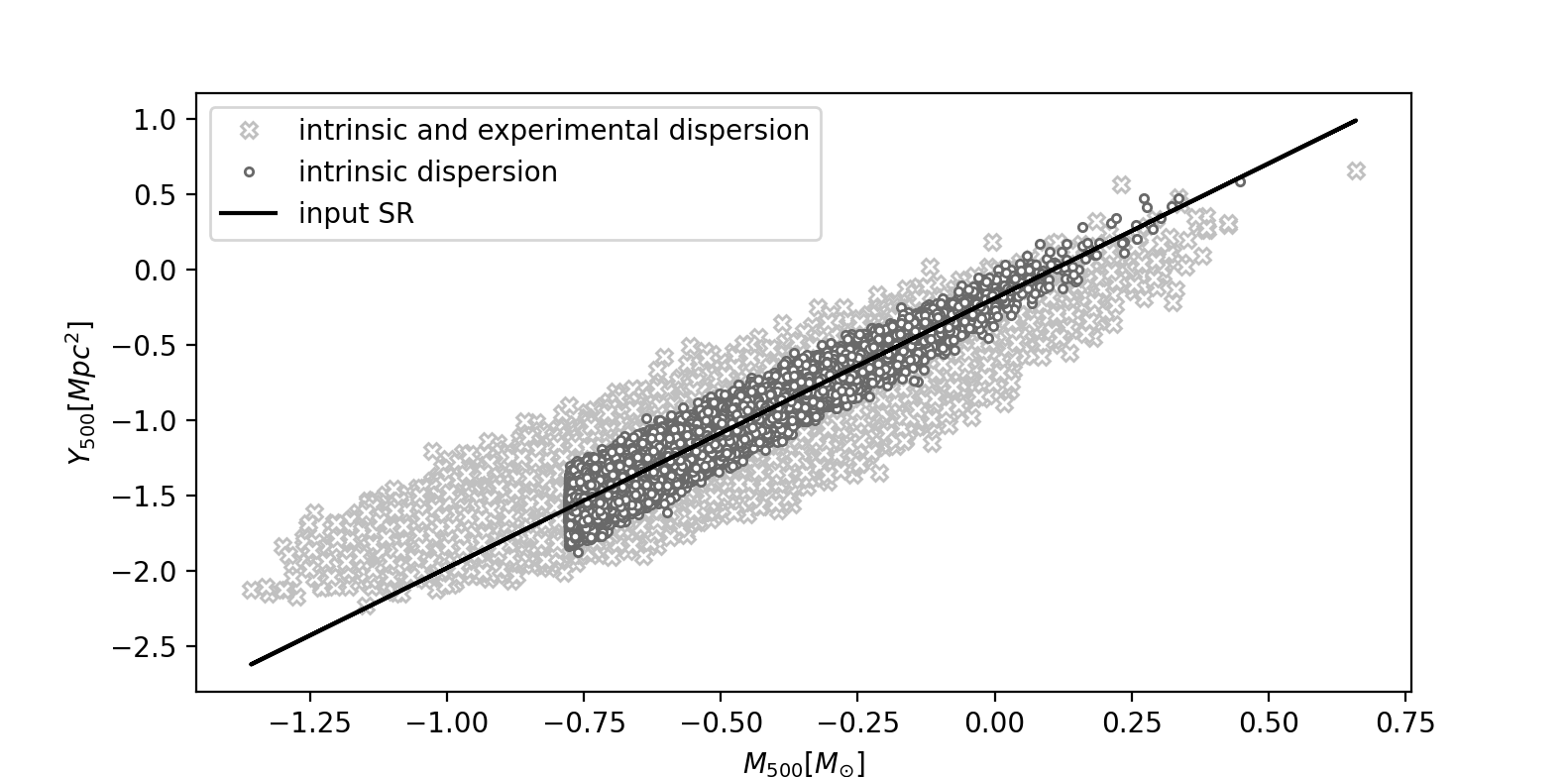}
    \caption{ Simulated clusters represented in the $Y$-$M$ plane. They all follow the same SR represented by the black line. Black circle points represent clusters affected by the intrinsic scatter only and grey cross points are the same clusters with an additional experimental dispersion.}
    \label{fig:scatter}
\end{figure}

The $Y$-$M$ scaling relation is derived assuming that all clusters are spherical, in hydrostatic equilibrium (HSE) and that the intra-cluster medium (ICM) is an ideal and isothermal gas.
These assumptions lead to a power law: 
\begin{equation}
    E(z)^{-\frac{2}{3}}\left(\dfrac{D_A^2 Y_{500}}{10^{-4} \mathrm{Mpc}^2}\right)=10^\alpha \left(\dfrac{M_{500}}{6\times 10^{14} \mathrm{M}_\odot}\right)^\beta
    \label{eq:YM_SR}
\end{equation}
with $E(z)=H(z)/H_0$ the dimensionless Hubble parameter.\\
However, the above-mentionned assumptions do not represent the reality of cluster population. This is the reason why there is an intrinsic scatter $\sigma$ with respect to the power law SR. 
In the end the $Y$-$M$ SR is defined as follows : 
\begin{equation}
    P(\log (Y_{{500}})|\log (M_{500}))=\mathcal{N}(\alpha + \beta \log (M_{500}),\sigma^2)
    \label{eq:real_SR}
\end{equation}
Thus the SR is defined with three parameters: $\alpha$ the intercept, $\beta$ the slope
and $\sigma$ the intrinsic scatter. Among these parameters, $\sigma$ is difficult to estimate as it is hidden behind the experimental dispersion as shown in figure \ref{fig:scatter}.
Indeed, the observed dispersion is a combination of the intrinsic and experimental ones. This is why, we use the LIRA software \cite{LIRA} to retrieve SR parameters and separate the two contributions to the total dispersion. LIRA is based on a MCMC Gibbs sampling algorithm and takes the variables, their errors and their correlations as inputs.\\
The NIKA2 collaboration \cite{NIKA2-general}  
is working with its SZ Large Program (LPSZ) \cite{LPSZ} on a new estimation of the $Y$-$M$ scaling relation and of the mean pressure profile \cite{H23}. 
They will be obtained from a sample of 38 clusters selected from Planck \cite{Planck} and ACT \cite{ACT} catalogs with redshifts between $0.5<z<0.9$. This sample has been observed with the NIKA2 camera at the IRAM 30-m telescope in the past years.  
The Planck estimation \cite{planck_YM} of the scaling relation was obtained with a sample of low redshift clusters ($z<0.45$), whereas the LPSZ one will be obtained with intermediate to high redshift clusters. Moreover, the NIKA2 angular resolution (17.6 arcsec resolution at 2 mm) \cite{NIKA2-performance} allows us to have access to cluster substructures and to study cluster dynamics. This will allow us to cover effects related to more disturbed clusters, including mergers and elliptical ones. \\
Before studying the effects corresponding to clusters physics, we must understand and take into account systematic effects from the selection function and from the data analysis (from maps to integrated quantities).\\ 
The LPSZ selection function has two distinct contributions. On the one hand, the selection functions of Planck and ACT, since LPSZ clusters have been selected from these catalogs. And on the other hand, the effect of the box selection, which is the main contributor to the LPSZ selection function. The latter was used to force the sample to be homogeneous and not influenced by the underlying cluster's mass distribution. Boxes are 2D bins in redshift and $Y_{500}$. However this box selection can not be trivially processed as there are 5 thresholds in the $Y_{500}$ axes and they cannot be treated individually. As shown by F. Kéruzoré \cite{these_FK}, it will induce 4 Malmquist-like biases, which results in a $\alpha$-bias in our sample.
\section{Systematic effect study}
\label{sec:sample_descr}
\begin{figure}
    %\centering
    \hspace*{-0.5cm}\includegraphics[scale=0.43]{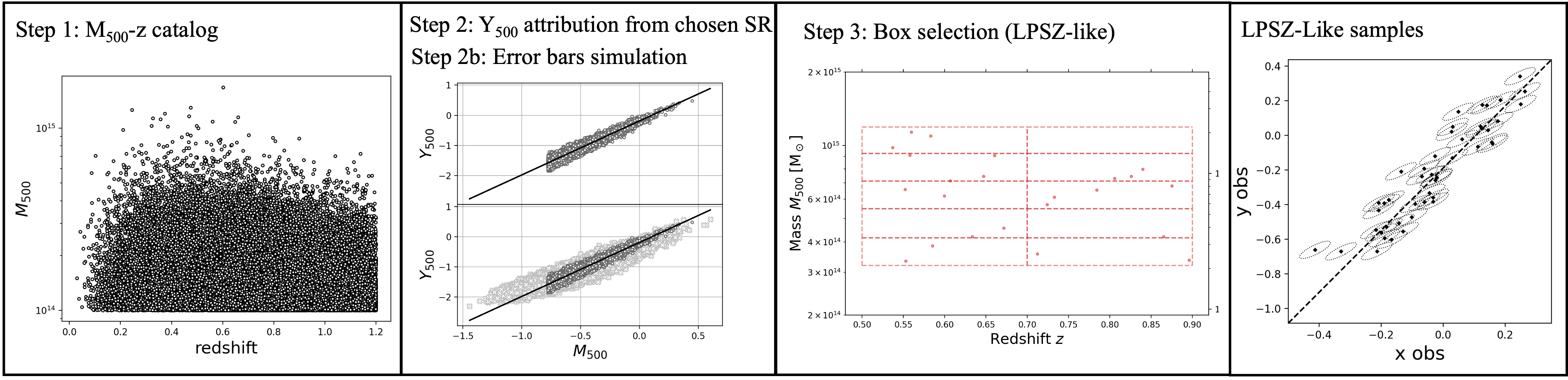}
    \caption{Steps followed to simulate a LPSZ-Like sample starting from a mass-redshift catalog. a $Y_{500}$ is given for each simulated clusters. Then the LPSZ selection function can be applied to this catalog.}
    \label{fig:sim_step}
\end{figure}

Systematic effects are studied with simulations. The goal of these simulations is to be as realistic as possible, but at the same time, to be simple enough so that we are able to control all parameters. The steps followed to simulate clusters are presented in figure \ref{fig:sim_step}.
In step 1, we simulate a cluster population with their associated masses and redshift following a mass function \cite{Tinker}%{\color{red}[ref]})
. In step 2, each cluster is given a $Y_{500}$ following a chosen SR, {\it i.e.} with known $\alpha$, $\beta$ and $\sigma$. Then, to simulate observations, each $Y_{500}$ and $M_{500}$ value is perturbed, {\it i.e.} a new value is drawn inside their error bars (Step 2b). This new value follows a 2D Gaussian distribution with mean the $Y_{500}$ and $M_{500}$ inputs. The expected errors and correlations of the panco2 ouputs are the covariance matrix of the distribution. In step 3, the LPSZ box selection is applied to this mock catalog which gives a realistic LPSZ-like sample with a known scaling relation. After this last step, the sample is regressed by LIRA \cite{LIRA}
thus obtaining the probability density distributions of the SR parameters.\\ 
Our goal is to study the impact of systematic effect on the estimation of SR parameters, so these steps are repeated 5000 times to avoid statistical effects. Each time, only the median value of the distributions estimated by LIRA is stored. Then, the 5000 values are compared to the input, in order to conclude on a possible bias. We repeat this study for different input SR. The Planck SR \cite{planck_YM} was taken as a reference, which means only one parameter of Planck SR was replaced each time by the values written in table \ref{tab:my_label}. The result of this study shows that, at this stage, the estimation of SR parameters are correlated with each other.\\
\begin{table}[h]
    \centering
    \small
    \begin{tabular}{|c|c|c|c|c|c|c|}
    \hline
        $\alpha$ & -0.53 & -0.33 & \textbf{-0.19} & -0.1 & -0.05 & \\
        \hline
        $\beta$ & 1 & 1.66 & \textbf{1.79} & 2.25 & 2.75 & 3\\
        \hline
        $\sigma$& 0 &0.025 & \textbf{0.075} & 0.1 & 0.12 & 0.15\\ 
        \hline
    \end{tabular}
    \caption{Fiducial values chosen for scaling relation parameters. Planck values \cite{planck_YM} are in bold.}
    \label{tab:my_label}
\end{table}
\begin{figure}[h]
    %\centering
    \hspace*{-1cm}\includegraphics[scale=0.58]{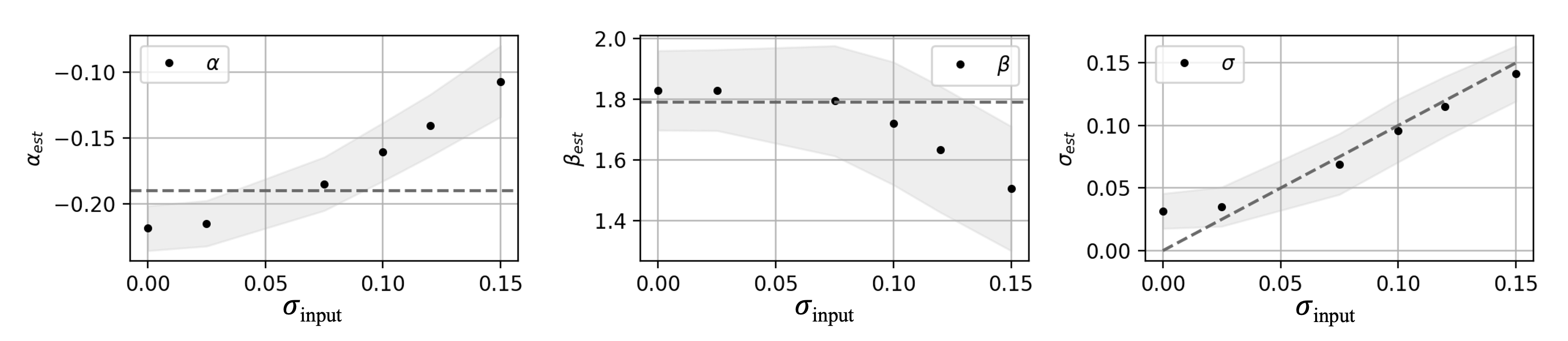}
    \caption{LIRA estimation of $\alpha$, $\beta$ and $\sigma$ as a function of different $\sigma$ input values. Black dashed lines represent input values. The points indicate the median of LIRA estimations and the shaded areas corresponds to the $68\,\%$ CL dispersion.}
    \label{fig:bias}
\end{figure}
In figure \ref{fig:bias}, we present the case for which $\sigma$ varies. For reasonable  values of $\beta$ (close to the $\beta_{Planck}$ estimation at the 2 $\sigma$ level) and all values of $\alpha$, we always retrieve input values with LIRA. As $\sigma$ increases, bias on $\alpha$ and $\beta$ increases linearly. This effect could be explained by the same mechanism as the Malmquist bias. On the right panel of figure \ref{fig:bias} right, LIRA always retrieves the input values of the scatter $\sigma$ without bias and with reasonable error bars, except for the two first points. For this sample size (around 5 clusters per box), below $\sigma \sim 0.025$ LIRA can not distinguish between the intrinsic and the experimental dispersions. Nevertheless, this should not be an issue for the LPSZ, since the value of $\sigma$ is expected to be much larger ($\sigma_{Planck}=0.075$).
Knowing that one parameter will always be retrieved, we can linearly parametrize biases as a function of $\sigma$ for $\alpha$ and $\beta$. To be more accurate, a power law parametrization has been done for $\alpha$ as a function of $\beta$ as there is a small bias for large values of $\beta$.  
\begin{figure}[h]
    \hspace*{-1cm}\includegraphics[scale=0.22]{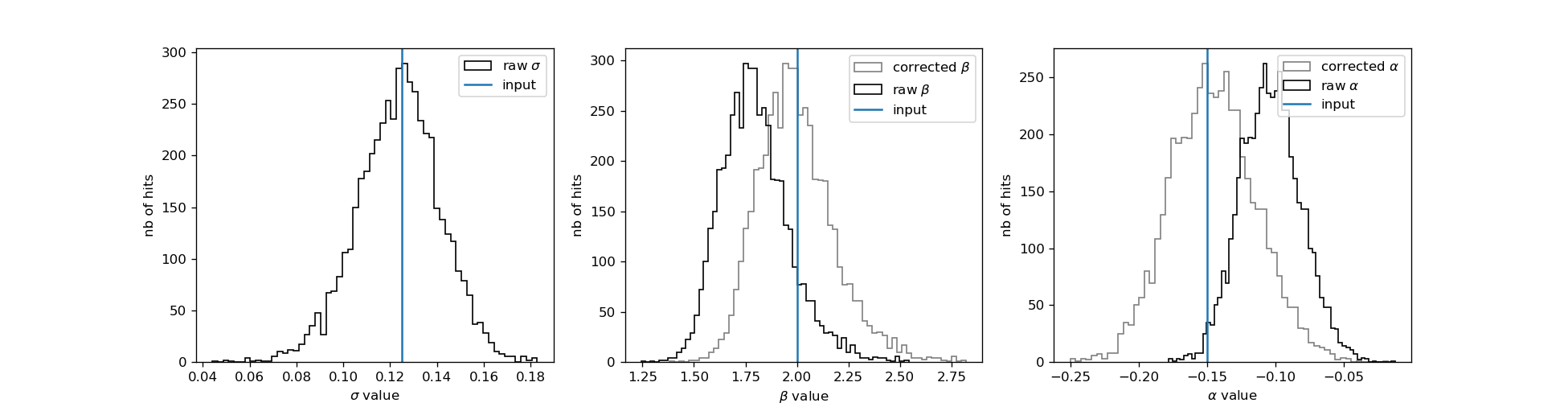}
    \caption{Distributions of the 5000 LIRA estimations of the SR parameters on samples described in section \ref{sec:sample_descr}. Black histograms represent the raw LIRA estimation, blue ones represent the estimation corrected with the parametrization. The blue lines represent the SR input values}
    \label{fig:corr_bias}
\end{figure}
%\vspace{-0.2}
In figure \ref{fig:corr_bias},
we test the bias parametrization for a given SR. On the left, we see that we do retrieve the input value of $\sigma$. For $\alpha$ and $\beta$ the raw estimations of LIRA are biased, but after applying the above mentioned correction, the final estimations are compatible with the input SR values. Hence, the box selection function has been accounted for. 
\section{Effect of the pipeline analysis}
Now that the selection function has been accounted for, we want to know what are the effects of the data analysis. For this purpose, we use the official LPSZ pipeline on simulated maps. 
We follow the steps described in section \ref{sec:sample_descr} (without error bar simulation). Then, maps are simulated assuming that clusters are spherical and relaxed. We assume a gNFW pressure profile \cite{gNFW} 
and typical white and correlated noises that were obtained from NIKA2-LPSZ data. We take into account other instrumental effects such as the NIKA2 beam and a typical transfer function \cite{NIKA2-performance}
. All these simulated maps has been given to panco2 \cite{Keruzore:2022tpj}  
which is the official software used by NIKA2-LPSZ to obtain the pressure profile and the integrated quantities from maps.
As a result, panco2 gives the probability density distributions of $Y_{500}$ and $M_{500}$ and their correlation. These are used as input values for LIRA that gives an estimation of SR parameters to be compared with the input values.\\
Two map samples were simulated, one with white noise and another one with correlated noise. Regarding the estimation of the integrated quantities with panco2, we always retrieve the input values within error bars in both cases. Note that for maps with correlated noise, there is a larger dispersion around input values: $2.5\%$ ($5\%$) for $Y_{500}$ ($M_{500}$). Whereas for white noise we have $1.1\%$ (2.2\%) for $Y_{500}$ ($M_{500}$).
With these two samples, we obtain two estimations of the SR parameters. We do retrieve the input SR within two sigma confidence level from both samples but with smaller error bars for the correlated noise case\footnote{This can be explained by the fact that we add a noise covariance matrix when we analyse maps with correlated noise.}.
\section{Conclusion}
Several systematic effects have been studied so far. Effects induced by the selection function can now be accounted for. Effects coming from data analysis have also been studied and no significant impacts have been observed. Understanding the effects due to the map analysis and the LPSZ selection function will help us to identify the impacts clusters physics have on the determination of the SR.
The next step is to study the impact of clusters deviating from the hydrostatic equilibrium, thus taking advantage of the NIKA2 high-angular resolution. 
All these investigations will be useful for the upcoming LPSZ SR.

% NIKA2 acknowledgements
\section*{Acknowledgements}
\small{We would like to thank the IRAM staff for their support during the observation campaigns. The NIKA2 dilution cryostat has been designed and built at the Institut N\'eel. In particular, we acknowledge the crucial contribution of the Cryogenics Group, and in particular Gregory Garde, Henri Rodenas, Jean-Paul Leggeri, Philippe Camus. This work has been partially funded by the Foundation Nanoscience Grenoble and the LabEx FOCUS ANR-11-LABX-0013. This work is supported by the French National Research Agency under the contracts "MKIDS", "NIKA" and ANR-15-CE31-0017 and in the framework of the "Investissements d’avenir” program (ANR-15-IDEX-02). This work has benefited from the support of the European Research Council Advanced Grant ORISTARS under the European Union's Seventh Framework Programme (Grant Agreement no. 291294). A. R. acknowledges financial support from the Italian Ministry of University and Research - Project Proposal CIR01$\_00010$. S. K. acknowledges support provided by the Hellenic Foundation for Research and Innovation (HFRI) under the 3rd Call for HFRI PhD Fellowships (Fellowship Number: 5357). 
}

\end{document}